\documentclass [a4paper,12pt] {article} 

\usepackage{amsmath}
\usepackage{amssymb} 
\usepackage{amsfonts}

\usepackage[square]{natbib}

\usepackage{graphicx}

\usepackage[pagebackref=true,colorlinks=true,citecolor=blue,linkcolor=blue,urlcolor = blue]{hyperref}

\title{\textbf{Error probability amplification} \\ \textbf{in cellular translation}}

\author{
	Vladimir N. Binhi \medskip
	\\ 
{\small	Prokhorov General Physics Institute of the Russian Academy of Sciences }\\ 
{\small 38 Vavilov St., 119991 Moscow, Russia }
            \smallskip 
	\\ 
 {  {\small \underline{vnbin@mail.ru} }
              }
       }

\begin{document}

\sloppypar 

\maketitle

\abstract{During cellular translation, incorporation errors occur. It is the addition of amino acid residues not corresponding to the  mRNA code. With an increase in the number of residues in the synthesized molecule, the probability of failure in at least one link increases rapidly, which leads to improper folding and loss of functionality of the entire molecule. A simple mathematical model is presented, which shows that the amplification factor equals approximately the number of links in the sinthesized sequence. We assume that the enzymatic processes of recognition of amino acids and their addition to the synthesized molecule include the formation of intermediate pairs of radicals with spin-correlated electrons. If a weak external magnetic field slightly changes the rate of quantum singlet-triplet conversion, then a significant change in the probability of occurrence of incorrect sequences of ribosomal translation occurs. } \\


\section{Introduction}

An extensive literature \citep {Barnothy-1964, Binhi-2011, Buchachenko-2014-e, Greenebaum-and-Barnes-2019} is devoted to the biological effects of weak magnetic fields (MFs). As is believed, MF can cause various, including toxic, effects in organisms \citep{Ghodbane-ea-2013}. The difficulty lies in the fact that there is no convincing physical mechanism that would provide a discernible shift in the probability of an individual act of a chemical reaction in the MF on the order of the geomagnetic field, about $50~\mu$T \citep{Binhi-and-Rubin-2022}. One of the mechanisms relates the biological effect of MFs to the presence of magnetic nanoparticles in organisms \citep{Kirschvink-ea-1985, Binhi-and-Chernavsky-2005}. However, some cell cultures and plants that respond to MFs do not contain magnetic nanoparticles. Therefore, search for a general molecular mechanism of the biological response to MFs continues.

The characteristic of the current state of the problem is that even the most plausible mechanism of magnetic biological effects, the Radical Pair Mechanism (RPM), provides only an insignificant, at best on the order of 0.01--0.1\%, response to a change in the MF of the geomagnetic field level \citep{Binhi-2002, Binhi-and-Prato-2018}.  Several orders of magnitude are missing to explain the facts of animal magnetic navigation \citep{Binhi-and-Rubin-2022}. The difficulty is that the coherence of spin quantum states, which provides the magnetic effect at physiological temperatures, takes place over a short time interval. In weak MFs, however, a noticeable effect occurs only at a sufficiently long coherence relaxation time. These conflicting trends prevent the occurrence of a magnetic RPM effect sufficient to explain the observations. The thermal relaxation time of electron spins in biological media at physiological temperature is about 1~ns in order of magnitude. Accordingly, noticeable magnetic effects in magnetochemistry usually arise in MFs exceeding 5~mT \citep{Steiner-ea-1989.CR}. To explain the biological effects of MFs of the geomagnetic field level of 0.05~mT, the thermal relaxation time of electron spins must exceed 100~ns, i.e., be comparable with the  Larmor precession period about 700~ns. It is unknown where and whether such states could appear in biological tissue. 

Apparently, the small primary changes that occur as a result of the action of MF must be somehow amplified to cause noticeable changes in concentrations of biochemical agents. This idea is not new. The enzymatic reaction, highly responsive to the enzyme concentration, was proposed in \citep{Weaver-ea-2000}. A chemical reaction in a mode close to bifurcation instability, when the reaction path can change dramatically with a slight variation in the reactants content, was proposed in \citep{Grundler-ea-1992, Riznichenko-ea-1994, Player-ea-2021}. However, these approaches have not been developed --- probably due to the impossibility of their experimental verification. The idea was put forward of a statistical amplification of parallel negligible magnetic signals from millions of photoreceptors by the brain. However, the gain has appeared to be insufficient to explain the magnetic navigation of animals based on the RPM \citep{Binhi-2023-e}. Moreover, this mechanism does not apply to nonspecific effects --- in the absence of evolutionarily developed magnetoreceptors.

In this paper, we pay attention to the fact that small initial signals can be statistically amplified, --- in the course of their accumulation in a long series of repeated events. It is known, for example, that a small physical effect of the effort of thought can become noticeable after millions of elementary acts of scattering in the quincunx or acts of generation of binary random events. The integrated deviation served as a tool for detecting the effect, --- as a tool that allowed one to accumulate minor regular deviations against the background of significant random variations \citep [see. e.g.] [p.\,319] {Binhi-2021-e}. In the same way, one would expect that magnetic effects, if they are due to primary minor signals under the action of an MF, could be reliably detected where they accumulate in a long sequence of elementary events.

The natural carrier of such an integrator in the cell is the process of gene expression. The essence of these processes is the multiple and almost error-free repetition of homogeneous acts of biochemical reactions involving biopolymers. Such cyclic processes are ideal integrators; they accumulate the probabilities of errors occurring in each step. Sooner or later, the accumulation of error probabilities leads to disruption of the functions of the synthesized and folding proteins, which one can record in the experiment. Therefore, a relatively stable magnetic effect is mainly expected where there are intensive processes of replication, transcription, translation, and folding, --- in cells under radiation or chemical stress.

As is known, there are almost no data on the sensitivity of pre-biological systems to a weak MF \textit{in vitro}. And if there are \citep{Novikov-ea-2020-e}, then their relationship with the effects of \textit{in vivo} remains questionable \citep{Hore-2012}. On the other hand, many works indicate that gene expression, which includes various processes of biopolymer synthesis, may be a prerequisite for the nonspecific effects of MFs. 

It is essential that cyclic processes of biopolymer synthesis are catalytic. They occur due to enzymes that directly produce  elongation of the biopolymer chain. Intermediate states of paired radicals with spin-correlated electrons can arise in the active sites of enzymes \citep [e.g.] [] {Afanasyeva-ea-2006, Buchachenko-2014-e}. Such quantum states are known to have magnetic sensitivity.
 
There are many studies of the magnetic effects on enzymatic activity in organisms. However they do not show any pattern that controls the occurrence of magnetic effect or links its magnitude with the parameters of magnetic exposure. The MF effect on enzymatic activity appears to be largely random. Apparently, the changes that occur at the molecular level --- when they occur --- depend significantly and ambiguously on many biochemical conditions, on the one hand, and also ambiguously affect the measured characteristics, on the other hand \citep{Binhi-2021-BEMS}.

In this work, we allow for the fact that observed nonspecific biological MF effects are random and caused by relatively weak MFs. The purpose of this work was to demonstrate a statistical model of the accumulation of small primary signals in the probability of local translation errors up to a level of more than a few percent. This mechanism would then explain observed effects of weak MFs without contradictions. We show that this model has the properties necessary to explain the specific magnetoreception and nonspecific magnetic response.

\section{Amplification of the incorporation errors in cellular translation}

The processes of biopolymer synthesis are diverse. These are DNA replication, transcription --- synthesis of complementary RNA, splicing, translation --- protein synthesis from amino acids under the mRNA code, and post-translational folding of the protein chain into a globule and its maturation. At each stage, random errors can occur, but with significantly different probabilities. There are perfect biochemical mechanisms for correcting replication and transcription errors; the likelihood of these errors is therefore tiny, on the order of $10^{-8}$ and $10^{-5}$, respectively, \citep{Mohler-and-Ibba-2017}. Translation errors occur much more frequently, with a probability of the order of $q=10^{-4}$--$10^{-3}$ per added amino acid \citep{Parker-1989, Kurland-1992}. Then about $1-(1-q)^{300} \sim 3$--26\% of synthesized molecules of 300 units contain at least one error. For such molecules, the chance of adopting a native conformation during folding significantly reduces. Often they are cytotoxic and cause harmful cellular effects \citep{Drummond-and-Wilke-2008}.  

Errors also occur at the folding stage --- one of the causes is the intricate geometry and topological nodes of the folding trajectories \citep{Nissley-ea-2022}. However, misfolding due to translation errors, as is believed, is more likely than due to the actual folding. Therefore, translation errors mainly control the accuracy of gene expression. Consequently, the possible influence of MFs on the probability of translation errors is the process where magnetic effects could manifest themselves at the biological level. However, the possibility of MF influence on translation errors, as far as we know, was not previously considered in theoretical models.

The translation is a complex multi-stage cyclic process that includes a variety of enzymatic reactions. The ribosome produces translation --- it is a macromolecular machine assembling amino acids into proteins. Below is a statistical model of ribosomal translation in which there is a low probability of a local incorporation, or substitution, error --- the appearance, in the synthesized chain, of a non-cognate amino acid residue that does not correspond to the mRNA blueprint. 

A simplification illustrating the occurrence of a noticeable translation error is as follows. Let a ribosome produce a protein chain of a large number $n$ of links with an equal probability $\xi$ of incorporation error. A native functional protein globule implies the absence of local errors at all $n$ links in the chain. Then $(1-\xi )^n$ is the probability of occurrence of error-free amino acid sequence, and
\begin{equation} \label{vererr}
p = 1- (1-\xi )^n \end{equation}  is the probability of the appearance of a defective molecule, i.e., the probability of a translation error. The sensitivity of $p$ to the local incorporation error probability $\xi $ is the derivative $ {\rm d}p/{\rm d}\xi = n(1-\xi )^{n-1}$. Its magnitude is maximum under the condition $n\approx (2\xi)^{-1} $ and can reach large values of $\sim (4\xi)^{-1} $ at small $\xi$.

The probability of correct translation of the entire molecule does not exceed unity. In particular, for $n\xi \sim 1$ and $n\gg 1$ we have $p\sim 1-1/e \approx 0.63$. This fact means that the result of changing the error probability when $\xi $ varies is not that the probability (\ref{vererr}) changes much, but that a change in the translation error $p$ by a few tenths occurs when varying very small $\xi $.

Most of proteins in the human body have a length of one to five hundred amino acids. The probability of a local failure is unlikely to reach $10^{-2}$ because almost all proteins would fold incorrectly otherwise. Therefore, the range $\xi < 10^{-3}$ is interesting.

\begin{figure}[htbp] 
	\centering
	\includegraphics[width=6cm]{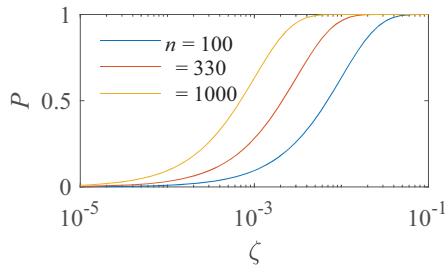}
	\parbox[b]{105mm}{ 
	\caption{\small The dependence of the probability of a translation error on the incorporation error probability $\zeta$ at different values of the length $n$ of a sinthesized protein.  
		\label{fig:01}}   }                            
\end{figure}

In a more realistic model, independent random variables $\xi_i$, $i=1,2,..,n$ on each link represent the failure probabilities. Let all of them have the same distribution with expectation ${\rm E}[\xi_i]= \zeta $ and variance ${\rm D}[\xi_i]=\sigma^2 $. In this case, the translation error 
\begin{equation*} \label{verfolrand}
p = 1- \prod _{i=1} ^n (1-\xi_i ) \end{equation*} becomes a random variable with mean $P={\rm E}[p]$ and variance $S^2= {\rm D}[p]$, respectively
\begin{equation} \label{MandS} {P} = 1- (1-\zeta )^n ,~~ S^2 =  \left[\nu +(1-\zeta )^2 \right]^n - (1-\zeta )^{2n} \end{equation} 
Function $P(\zeta,n)$ is shown in Fig.~\ref{fig:01}; it demonstartes that significant values of $P$ can occur even at minor values of $\zeta $.

Expanding (\ref{MandS}) into a series in $\zeta n$, one can see that in the region of large $n$ and small $\zeta < 1/n$ the relations 
\begin{equation} \label{finalres} { P} \approx \zeta n \end{equation} and $S \approx \sigma\sqrt{n}$ take place. The average incorporation error $\zeta$ increases by a factor of $n$, and the fractional value $S/{P}$ of variations decreases in proportion to $\sqrt{n}$. Since the length of most proteins is about a hundred and more, the main amplification effects arise from the change in the average value of the errors and not from their variations. 

Since small probabilities of substitution errors lead to significant variations in the translation error, it makes sense to assume that MF can change those probabilities of local errors. Then, a magnetic effect, even being minor initially, could manifest itself in significant changes in the concentration of nonfunctional proteins. These, in turn, would lead to an additional load on biochemical adaptation mechanisms and the appearance of noticeable biological effects.

Cyclic processes of biopolymer synthesis are those produced by enzymes that elongate the biopolymer chain. In translation, the ribosome cyclically reads information from mRNA and attaches a suitable amino acid to the synthesized protein chain. We suggest that this enzymatic recognition/attachment in the ribosomal active site includes the electron transfer and formation of the intermediate radicals in the singlet state. If a necessary amino acid enters the active site, then the process ends with chain elongation without error. If a non-cognate amino acid gets into the active site, this amino acid is not accepted. The radical pair decomposes into the initial state of the enzyme and amino acid, and the latter goes out into the cytoplasm. An external MF causes a singlet-triplet (S-T) conversion and, thereby, reduces the probability of decomposition and, accordingly, there occurs a possibility of including an incorrect amino acid group in the protein chain. A local incorporation error appears. Thus, MF increases the likelihood of the appearance of defective protein chains that cannot further acquire the correct conformation or become functional globules. 

This scenario mixes many processes before and in the course of the ezymatic chain elongation and should be considered an idealized illustration of possible biochemical kinetics. This scenario, while simplified in terms of the incorporation error's dependence on the MF, neatly describes how the translation error (\ref{finalres}) accumulates.

\section{Discussion}

First, we emphasize that formula (\ref{MandS}) describes the accumulation of probabilities, not physical changes. Physical changes happen suddenly at a random moment in translation. That transformation of small probabilities into a significant one does not occur in time as the chain elongates. It refers to the result of the synthesis, --- an entire molecule.

Until now, to explain biological magnetic effects on the basis of the RPM, one had to assume a long thermal relaxation time of electron spins in biological tissues, on the order of 100~ns and more \citep{Hore-and-Mouritsen-2016}. This was required in order to reconcile the theoretical calculation with the observed magnitude of magnetic effects in magnetic navigation of animals and in laboratory magnetobiology of non-specific effects, where the magnitude reaches up to about 10\% while the RPM magnitude is about 0.1\% in the geomagnetic field and is approximately proportional to the thermal relaxation time. The presented model of a statistical amplification of weak primary magnetic signals is free from this shortcoming. The translational statistical amplification explains the missed two orders of magnitude at the most probable thermal relaxation time of 1~ns.

As is known, strong MFs, of the order of 1~T and more, used in magnetic tomography, are safe at limited times of MF action of the order of 20--30 min. This no harmful effects, of course, does not mean the absence of any biological effects of such an MF in general. The above-presented regularities explain why strong MFs exceeding the geomagnetic field by four or more orders of magnitude do not lead to likewise strong magnetic effects in comparison with those in the MF of the geomagnetic level. Due to the kinetic limitations of the RPM, the magnetic effect is saturated already at the molecular level. 

In general, it is difficult to predict how the deviation in the level of translation errors from its natural level in the geomagnetic field will affect the observed biological characteristics. It is unclear what is the subsequent transduction pathway of the mistranslation signal to a measured reaction. Even the sign ($\pm$) of the body's response to MF is not clear. For example, an increase in the concentration of reactive oxygen species, in response to an MF change, as a secondary biochemical effect could occur with both an increase and a decrease in MF relative to the natural level to which the body is adapted. Therefore, when discussing the connection between the aberrant translation and the values measured in the experiment, it would be reasonable not to pay attention to the sign of effects but to interpret only the qualitative features of the MF response.

Model validation would be possible not only \textit{in vivo}, but also \textit{in vitro} --- in laboratory biochemical translation systems. They are being actively developed \citep{Chicherin-ea-2020}. The dependence of the reaction yield on the MF proves the existence of an intermediate {S-T} state of a pair of electrons in magnetochemistry. Similarly, observation of the MF-dependent concentration of incorrectly synthesized molecules \textit{in vitro} would be a direct evidence of the existence of an intermediate {S-T} state of electron pairs during translation.

In calculations \citep{Hore-and-Mouritsen-2016}, the relative RPM magnetic effect in a simple configuration ``two electrons, one proton'' was about 10\% when the MF changed by the value of the geomagnetic field. At first glance, this is enough to explain the nonspecific effects of the geomagnetic field. The problem, however, is that a long thermal relaxation time of electron spins, 1~$\mu$s, was used in the calculations. The authors of \citep{Kattnig-ea-2016, Worster-ea-2016} attempted to substantiate such a great value theoretically. However, there are no experiments so far that would confirm the existence of this long relaxation time, with the exception of exotic systems such as fullerenes, which have nothing to do with biology. Reliable evidence of the spin relaxation time in radical pairs could be given by measurements of the EPR linewidth in the geomagnetic field or by measurements of the spin magnetic effects \textit{in vitro} and their comparison with calculations. To the best of our knowledge, EPR signals in the geomagnetic field from the electron spins of transient pairs of radicals have not been observed in biochemical reactions. At the same time, the experience of spin chemistry, which agrees with the theory, indicates a faster spin relaxation under the conditions under discussion, about 1~ns. Perhaps, at the most, 10~ns. The magnetic effects fall in the same way, by a thousand or a hundred times, since they are roughly proportional to the spin relaxation time --- unless the rate of chemical process is a ``bottleneck'' suppressing magnetic effects to an even greater extent. The correct value of the RPM effect is thus only 0.01--0.1\% per 50~$\mu$T --- a figure that follows from the fundamental relation $\gamma H \tau \sim 1$ \citep{Binhi-and-Prato-2018}.  

Even if the spin relaxation time in radicals is about 1~$\mu$s, and even more so if it is noticeably less, the RPM cannot explain the specific sensitivity of some seasonally migratory species to geomagnetic variations in the MF at the level of tens of nT. There is also no explanation for often observed nonspecific effects of the geomagnetic storms on the state of organisms \citep [e.g.] []{Pishchalnikov-ea-2019}. This is a disappointing situation --- we cannot explain the magnetic biological effects without appeal to some obscure tricks of natural biological evolution.

What new can the above-described statistical mechanism bring to this state of affairs? First, the statistical amplification of initially weak magnetic RPM signals in the process of cellular translation makes it possible to raise the effect to the level of 5--50\%. Such magnitudes are observable and verifiable. 

It is essential that the statistical mechanism indicates a definite space location in the biochemical machinery where the primary magnetic biological effects occur. This region is the active site of the tRNA--ribosome complex, where enzymatic processes of the recognition of cognate amino acids and their attachment to the growing protein chain take place. Thus, the statistical mechanism can also be validated by biochemical methods, --- in addition to testing the qualitative features that follow from the RPM. Finally, we note that the magnetic influence on such a general molecular machine as the ribosome, which synthesizes many different proteins, is consistent with the experimentally observed fact that non-specific magnetic biological effects are mostly random effects \citep{Binhi-2021-BEMS}, not allowing simple averaging.

In general, the amplification of the probability of local errors is valid for any process, the result of which would be error-free only if there were no errors at each step in a long series. In addition to translation, these could be replication and transcription. As is known, cells have developed surprisingly perfect mechanisms for repairing replication and transcription errors. However, errors still occur at individual steps. Their low probability can accumulate and lead to cellular stress. It is not yet known whether an MF can influence the probability of these errors.

\section{Conclusion}

The statistical mechanism is related to non-specific effects that have been defined as not provided by specialized MF receptors and are thus an epiphenomenon of the vital functions of organisms in the Earth's MF. Perhaps it is also related to the specific effects of magnetoreception in some animals. The statistical mechanism means an increase in the probability of errors in biopolymers in MFs --- due to the amplification of the local errors probability --- and, consequently, the appearance of a significant fraction of nonfunctional proteins consisting of hundreds or more amino acid units. The amount of amplification is approximately proportional to the length of the synthesized biopolymer sequence.

In nonspecific effects of the biological response to a weak MF, the primary MF targets are most likely spin-correlated pairs of radicals with a typical thermal spin relaxation time of the order of 1~ns. To explain such effects, it is not necessary to assume that the relaxation time reaches 100~ns or more, --- an assumption that is most likely incorrect, but often assumed by default to match theoretical models with experiment.

The mechanism could be experimentally verified by testing those qualitative features that coincide with the RPM features. These are independence from the direction of the external MF; saturation of the magnetic effect with both increasing and decreasing MF from the geomagnetic level on a log scale; independence from the MF frequency in the low-frequency range, if this frequency is not that of any biological rhythm; otherwise, a frequency-dependent response unrelated to the RPM may appear.

In addition, the statistical mechanism indicates the active center of a ribosome as the primary site, where magnetic effects occur, creating the possibility of verification by biochemical methods;  predicts a random nature of the observed response to a weak MF; explains biological response to a hypomagnetic field; and agrees with the diversity of cell responses to a weak MF.

\vspace{6pt} 

The author declares no conflict of interest.  

\medskip This research was funded by the Russian Science Foundation grant No. 22-22-00951, {https://rscf.ru/en/project/22-22-00951/}


\end{document}